\documentclass[a4paper, twoside, twocolumn, prl]{revtex4} 

\usepackage{graphicx}
\usepackage{textcomp}		
\usepackage{longtable}

\ifx\pdfoutput\usepackage[pdftex]{hyperref} 
\else\usepackage{hyperref}	
\fi

\begin{document}

\title{Fracture strength and Young's modulus of ZnO nanowires}

\author{S.~Hoffmann, F.~\"{O}stlund, J.~Michler}
\affiliation{EMPA Materials Science and Technology, Feuerwerkerstrasse 36, CH-3602 Thun, Switzerland}
\author{H.~J.~Fan, M.~Zacharias}
\affiliation{Max Planck Institute of Microstructure Physics, Weinberg 2, D-06120 Halle, Germany}
\author{S.~H.~Christiansen}
\affiliation{Martin-Luther-University Halle-Wittenberg, Hoher Weg 8, D-06109 Halle, Germany}
\affiliation{Max Planck Institute of Microstructure Physics, Weinberg 2, D-06120 Halle, Germany}
\author{C.~Ballif}
\affiliation{Institute of Microtechnology, University of Neuch\^{a}tel, A.-L. Breguet 2, CH-2000 Neuch\^{a}tel, Switzerland}

\begin{abstract}
The fracture strength of ZnO nanowires vertically grown on sapphire substrates was measured in tensile and bending experiments. Nanowires with diameters between 60 and 310 nm and a typical length of 2 \textmu m were manipulated with an atomic force microscopy tip mounted on a nanomanipulator inside a scanning electron microscope. The fracture strain of ($7.7 \pm 0.8$)\% measured in the bending test was found close to the theoretical limit of 10\% and revealed a strength about twice as high as in the tensile test. From the tensile experiments the Young's modulus could be measured to be within 30\% of that of bulk ZnO, contrary to the lower values found in literature.
\end{abstract}

\maketitle

\section{Introduction}
Research interest in semiconductor nanowires (NWs) has increased exponentially over the past few years, driven by the NWs potential to act as key components in future integrated circuits, optical and nano electro mechanical systems (NEMS). There are a number of recent reviews about fabrication \cite{fan2006_700}, and their assembly for applications of semiconductor NWs in electronics \cite{thelander2006, li2006} and photonics \cite{li2006, pauzauskie2006}. In the particular case of ZnO NWs the main research focus has been on their optical \cite{djurisic2006} and electronic properties \cite{fan_zy2004,konenkamp2005} because of their wide and tunable bandgap, high room temperature exciton binding energy, and surface sensitivity to environments. Only little attention has been given to their mechanical properties, although they are important to the application of NWs as mechanical devices such as actuators and atomic force microscope (AFM) tips.

The Young's modulus of ZnO NWs or nanobelts has been measured to be $\approx$58 GPa by a mechanical resonance experiment \cite{huang2006}, $\approx$52 GPa by dual-mode resonance \cite{bai2003}, (31$\pm$2) GPa by a 3-point bending test with an AFM \cite{ni2006} and (29$\pm$8) GPa by a single clamped NW bending experiment with an AFM \cite{song2005}. Only Chen {\em et al.} observed a size dependence of Young's modulus \cite{chen2006}. With a resonance experiment they measured 140 GPa for NWs with diameters larger than 200 nm and up to 220 GPa for NWs with a diameter down to 50 nm. All of these experiments measured the Young's modulus by bending NWs, which is generally referred to as the bending modulus.

With a nanomanipulation robot arm mounted inside a scanning electron microscope (SEM) we performed both bending and tensile experiments on ZnO NWs. We also present results on the Young's modulus determined from a tensile test, which has a more homogeneous stress distribution than the bending configuration and the advantage that the result can be compared directly with the value for bulk ZnO (144 GPa along [0001], computed from elastic constants from \cite{wern2004}).

\section{Elastic beam theory}
No plastic deformation was observed deflecting the NWs. Even strongly deflected NWs (deflection/length $>$ 0.3) returned back to their original position when released. All strain measured in this work can thus be assumed to be purely elastic.

Consequently, in the bending experiment, the strain in the NW can be calculated with elastic beam theory. The strain depends on the diameter $d$, the length $l$, and the deflection $s$ of the NW. The maximum strain induced in the NW is located at its root where it is attached to the substrate and is 
\begin{equation} \label{bendstrain}
  \epsilon_{max}= \frac{3}{2}\frac{d}{l^2}\ s\ .
\end{equation}
For the geometries faced with in the present experiments, the strain calculated by Eq. \ref{bendstrain} differs by only $\pm$10\% from the first principal strain calculated by finite element analysis \cite{hoffmann2006}.

In a uniaxial tensile test, the stress induced in the NW is uniform and given by $\sigma=F/A$, where $F$ is the force applied to the NW and $A$ is its cross section. For a NW of diameter $d$ the stress is
\begin{equation} \label{tensstress}
  \sigma = \frac{4F}{\pi \ d^2} \ .
\end{equation}

With the initial length $l_0$ and the length $l$ of the strained NW, the Young's modulus $E$ can be calculated by measuring the strain $\epsilon = (l-l_0)/l_0$, 
\begin{equation} \label{young}
  E = \frac{\sigma}{\epsilon} = \frac{l_0}{l-l_0}\  \sigma \ .
\end{equation}

\section{Growth process of the ZnO NWs}
Single crystal ZnO NWs were synthesized via thermal evaporation and deposition inside a horizontal split quartz tube furnace (Carbolite HST 12/400). An alumina boat loaded with ZnO and graphite powder mixture (1:1 weight ratio) was located at the center of the third heating zone. The substrates, $a$-plane oriented $\alpha$-Al$_2$O$_3$ sapphire single crystals, were coated by a thin sputtered 3 nm Au film. The substrate was then placed above the source boat. The reaction tube was heated up by 20 \textdegree C/min to the desired temperatures and cooled down naturally to room temperature after the experiments. For samples A, B, and C the temperatures were 850, 850, and 820 \textdegree C respectively and the corresponding dwelling times were 3, 2.5, and 2.5 hr. The chamber pressure was maintained at 200 mbar by a constant flow of Ar gas and pumping. The residual air in the chamber provided the oxygen.

\begin{figure} 
    \includegraphics[width=0.45 \textwidth]{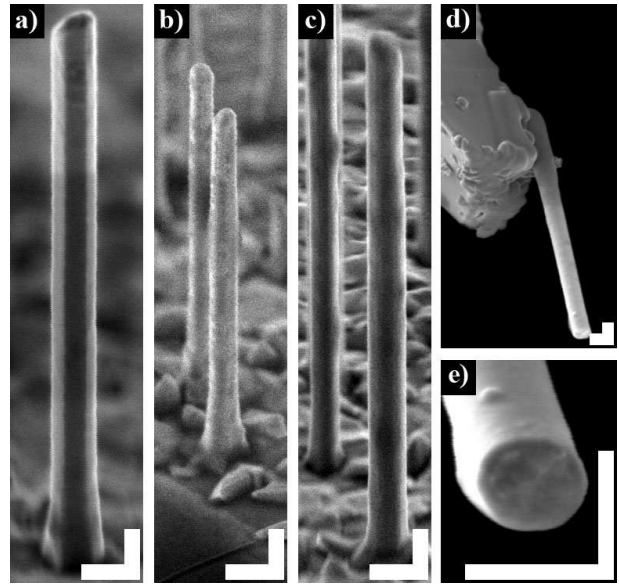}
    \caption{\label{hr_images}
    Field emitter SEM pictures of NWs. {a) to c): NWs of samples A, B and C respectively. d) NW attached to an AFM tip. It does not need a thick layer of carbonous deposit to make a firm bond. e) Fracture surface of a NW from a tensile test. For all the NWs that did not shatter the fracture plane was (0001). All scalebars correspond to \mbox{250 nm}, because of the tilt angle the vertical direction is contracted.}}
\end{figure}

Under our synthesis conditions it is most likely that both vapor-liquid-solid and vapor-solid mechanisms played a role for the growth. On one hand, the molten gold provided the necessary nucleation sites for Zn/ZnO vapors, leading to a vapor-liquid-solid growth process; on the other hand, vapor-solid can be a dominant growth process at temperatures in the range of 820-850 \textdegree C, giving rise to widening of the diameters via a lateral growth \cite{fan2006_561}. The density of NWs on the substrate is relatively low, which is convenient for single NW manipulation by the AFM tip. A NW of each sample is shown in Fig. \ref{hr_images}. A low magnification overview of the NWs (sample C) and their luminescence properties can be found in a previous work \cite{fan2005_023113}.

\section{Experimental setup and manipulation}
An AFM tip (AdvanceTEC, 3 and 45 N/m, NanoWorld Group) was fixed to a piezoelectric slip-stick robot arm (MM3A, Kleindiek Nanotechnik) with two rotational and one linear axis. The substrate with the NWs was mounted on an $xyz$ piezo stack (P-620.2CD and P-62.ZCL, Physik Instrumente) with 50 \textmu m range and sub-nanometer resolution. The whole setup was inside a SEM (S-3600N, Hitachi Science Systems) with the NWs at an angle of 60\textdegree \space to the electron beam.  With the SEM stage the NW of interest could be moved into the field of view. The coarse positioning of the AFM tip toward the sample was done with the robot arm and the fine positioning, as well as the manipulation of the NWs, was achieved by moving the sample with the piezo stack.

\begin{figure}
    \includegraphics[width=0.45 \textwidth]{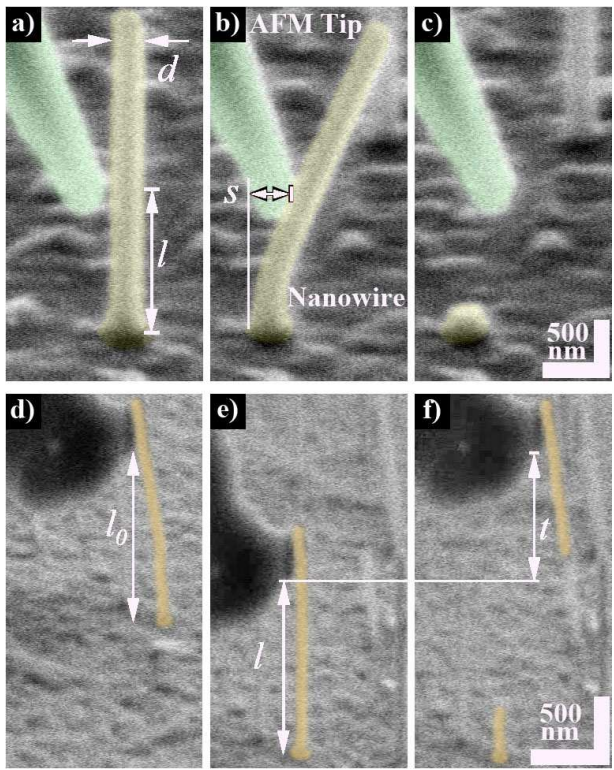}
    \caption{\label{experiment}
    {(Color online) a) - c): Sequence of a bending experiment. b) is the last image before fracture from which the deflection s can be read out. d) - f): Sequence of a tensile experiment. Note the precurved shape of the NW in d), and that it is straight just before fracture in e). In f) the AFM tip snapped back to the original position it had in d), which shows that no vertical stress was exerted on the NW at the beginning of the experiment. This is important for the measurement of Young's modulus.}}
\end{figure}

In the bending experiment the NWs were bent perpendicularly to the electron beam, so that the deflection $s$ could be read out directly from the SEM image. By applying the force in the middle along the length of the NW (Fig. \ref{experiment}a) the NW could effectively be shortened in order not to get too large deflections, for which Eq. \ref{bendstrain} would not hold anymore \cite{hoffmann2006}. From the last image before fracture the length $l$, diameter $d$, and deflection $s$ were extracted (Fig. \ref{experiment}b). With these parameters the maximum strain was calculated by Eq. \ref{bendstrain}.

In the tensile experiment the AFM tip was first brought into mechanical contact with the top of the NW. By scanning the electron beam for a couple of minutes over the contact region, carbonaceous contaminants originating from the sample surface and the residual gas of the SEM chamber were deposited in the interface. This formed a joint that was stronger than the NW itself and allowed us to pull on the NW until it fractured (Fig. \ref{hr_images}d). While pulling on the NW using the piezo stack, the SEM images were recorded to a video file. From the back leap $t$ of the cantilever at fracture and its spring constant $k$ the applied force at fracture $F=kt$ could be calculated (Fig. \ref{experiment}f). Prior to the experiment the NWs were imaged with a field emission SEM (S4800, Hitachi Science Systems) to measure their diameter. The fracture stress of the NW was then calculated by Eq. \ref{tensstress}. During retraction of the NW from the AFM tip the cantilever bent and caused an undesired deflection of the nanowire. This could partly be compensated for by moving the NW laterally, but because of the rigidity of the AFM tip-NW bond the NWs still were bent. To minimize the curvature at failure the NWs were prebent in the opposite direction before they were firmly attached to the AFM tip by pulling on them horizontally (Fig. \ref{experiment}d).

To calculate Young's modulus from the tensile test the critical parameter to extract was the length difference $l-l_0$ because of the limited resolution of the SEM. A homemade program based on a cross correlation algorithm located the position of the tip-NW joint and the root of the NW in each image of the video file. Despite of the noise present in the images the accuracy still was $\pm 1$ pixel. The length difference $l-l_0$ typically spanned 10 pixels so the strain could be determined within $\pm 10$\%. With the stress measured just prior to failure the Young's modulus was calculated according to Eq. \ref{young}.

\section{Results \& discussion}

Because of the similarity of their growth conditions the sample A, B, and C are indistinguishable from one another in terms of overall size, crystal structure (wurtzite, growth along [0001]), and luminescent properties (similar to those in Ref. \cite{fan2005_023113}). The only difference that could matter for this study is the crystalline quality, i.e. the defect density, which depends on the local Zn and O vapor concentrations during growth.

\begin{figure}
    \includegraphics[width=0.45 \textwidth]{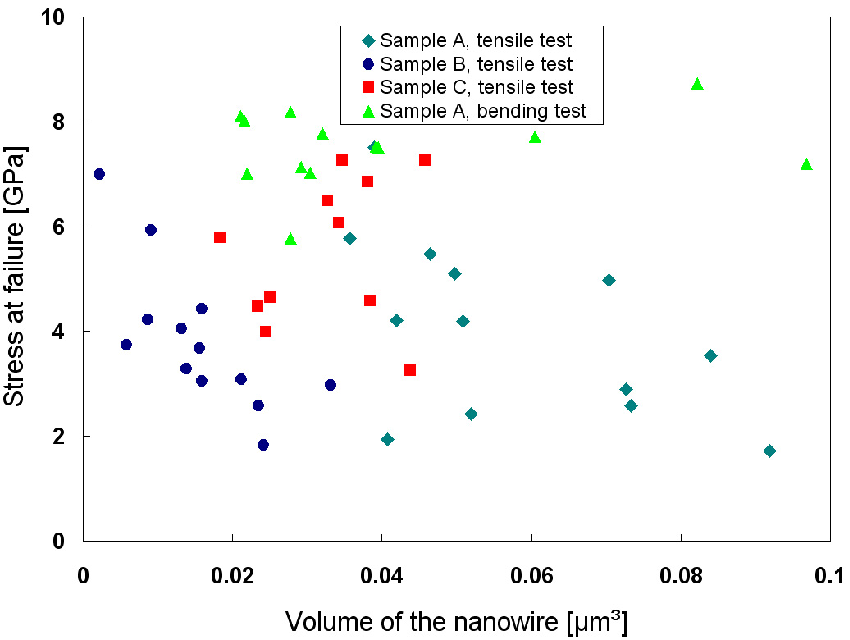}
    \caption{\label{stressvol}
    {(Color online) The maximum stress at failure plotted against the volume of the nanowires.}}
\end{figure}

Fracture strength of brittle materials is known to depend strongly on sample size \cite{namazu2000}, in particular the volume. In the case of NWs with a high surface to volume ratio, however, it has not yet been shown that the volume of the NW is the most relevant parameter. The overall geometry of all 50 experiments is therefore included in Table \ref{restable}. The average maximum strain measured in the bending experiment was (7.7$\pm$0.8) \% for sample A. The average tensile fracture strength was (4.0$\pm$1.7) GPa, (3.7$\pm$1.3) GPa and (5.5$\pm$1.4) GPa for sample A, B and C respectively. The measured Young's modulus for sample C is 97$\pm$18 GPa. The tensile and bending strengths are plotted against the NW volume in Fig. \ref{stressvol}. No clear size dependence can be observed, neither with respect to the volume, nor to the diameter, length or volume/surface ratio (not shown here). To calculate the stress in the bending experiments the Young's modulus measured from the tensile experiment, 97 GPa, was used.

\begin{figure*} 
\begingroup
\squeezetable 
\begin{longtable*}{ccccc|cccc|cccc|ccccccc} 
   \caption{ \label{restable}
   {The results of the bending and tensile tests as well as the measured Young's moduli with the dimensions of all the NWs tested. To calculate the stress in the bending experiment, the measured Young's modulus of 97 GPa was used.}} \\ \hline \hline
	  \multicolumn{5}{c|}{Bending experiment} & \multicolumn{15}{c}{Tensile experiment} \\
	  \multicolumn{5}{c|}{Sample A}&\multicolumn{4}{c|}{Sample A}&\multicolumn{4}{c|}{Sample B}&\multicolumn{7}{c}{Sample C} \\ \hline
	  $l$ & $d$ & $s$ & $\epsilon$ & $\sigma$ & $l_0$ & $d$ & $t$ & $\sigma$ & $l_0$ & $d$ & $t$ & $\sigma$ & $l_0$ & $l-l_0$ & $d$ & $t$ & $\epsilon$ & $\sigma$ & $E$  \\
	  {[\textmu m]} & {[\textmu m]} & {[\textmu m]} & {[\%]} & {[GPa]} & {[\textmu m]} & {[\textmu m]} & {[\textmu m]} & {[GPa]} & {[\textmu m]} & {[\textmu m]} & {[\textmu m]} & {[GPa]} & {[\textmu m]} & {[\textmu m]} & {[\textmu m]} & {[\textmu m]} & {[\%]} & {[GPa]} & {[GPa]}  \\ \hline
      .98 & .19 & .20 & 5.9 & 5.8 & 2.0 & .24 & 1.8 & 1.7 & 2.0 & .13 & .54 & 1.8 & 3.1 & .10 & .14 & 1.1 & 3.3 & 3.3 & 100 \\
      .98 & .17 & .27 & 7.2 & 7.0 & 1.4 & .20 & 1.4 & 1.9 & 2.3 & .11 & .63 & 2.5 & 3.1 & .13 & .10 & .74 & 4.2 & 4.0 & 94  \\
      .93 & .20 & .20 & 7.2 & 7.0 & 2.2 & .17 & 1.3 & 2.4 & 2.1 & .14 & 1.1 & 2.9 & 2.6 & .11 & .11 & .95 & 4.3 & 4.5 & 104 \\
      1.0 & .19 & .27 & 7.3 & 7.1 & 2.0 & .22 & 2.2 & 2.6 & 2.0 & .10 & .58 & 3.0 & 2.4 & .11 & .14 & 1.7 & 4.6 & 4.6 & 100 \\
      2.3 & .23 & 1.2 & 7.4 & 7.2 & 2.2 & .21 & 2.3 & 2.9 & 1.7 & .13 & .92 & 3.0 & 2.8 & .16 & .11 & .95 & 5.6 & 4.7 & 83  \\
      1.1 & .22 & .27 & 7.7 & 7.5 & 2.6 & .20 & 2.7 & 3.5 & 1.2 & .12 & .90 & 3.2 & 2.4 & .20 & .10 & 1.0 & 8.0 & 5.8 & 72  \\
      1.1 & .21 & .31 & 7.7 & 7.5 & 1.9 & .18 & 2.6 & 4.2 & 1.5 & .11 & .91 & 3.6 & 3.5 & NA  & .11 & 1.4 &     & 6.1 &     \\
      1.6 & .22 & .64 & 7.9 & 7.7 & 2.4 & .15 & 1.7 & 4.2 & 1.3 &.076 & .41 & 3.6 & 2.5 & .13 & .13 & 2.0 & 5.0 & 6.5 & 130 \\
      1.1 & .19 & .33 & 8.0 & 7.8 & 2.2 & .20 & 3.6 & 5.0 & 1.5 & .11 & .87 & 4.0 & 2.8 & NA  & .13 & 2.1 &     & 6.9 &     \\
      .82 & .18 & .20 & 8.2 & 8.0 & 2.7 & .15 & 2.1 & 5.1 & 2.1 &.073 & .43 & 4.1 & 2.6 & NA  & .15 & 3.0 &     & 7.3 &     \\
      .80 & .18 & .19 & 8.3 & 8.1 & 2.3 & .16 & 2.5 & 5.5 & 2.0 & .10 & .84 & 4.3 & 2.6 & NA  & .13 & 2.2 &     & 7.3 &     \\
      1.1 & .18 & .37 & 8.4 & 8.2 & 2.6 & .13 & 1.8 & 5.8 & 1.4 &.090 & .92 & 5.8 &     &     &     &     &     &     &     \\
      1.1 & .31 & .23 & 9.0 & 8.7 & 2.0 & .16 & 3.5 & 7.5 & .77 &.060 & .48 & 6.8 &     &     &     &     &     &     &     \\ \hline
      \multicolumn{3}{r}{Average $\epsilon$, $\sigma$:} & 7.7$\pm$0.8 & 7.5$\pm$0.7 & \multicolumn{3}{r}{Average $\sigma$:\ }& 4.0$\pm1.7$ & \multicolumn{3}{r}{Average $\sigma$:\ }& 3.7$\pm$1.3 & \multicolumn{4}{r}{Average $\epsilon,\ \sigma, \ E$:} & 5.0$\pm$1.5 & 5.5$\pm$1.4 & 97$\pm$18 \\
      \hline \hline
\end{longtable*}
\endgroup
\end{figure*}

In the tensile test the NWs either shattered or broke into two pieces at failure. Those that shattered showed a higher strength than those that broke into two pieces, probably because of the lack of a dominant defect that otherwise would have lead to an early failure at a specific location. At high stresses, more elastic energy is stored in the NW. The liberation of this elastic energy could be the cause for shattering of the highly stressed NWs. For the NWs that broke into two pieces the fracture surface was always a (0001) plane (Fig. \ref{hr_images}e), the cleavage plane of ZnO. In spite of the care taken to avoid a lateral deflection of the NW at failure, deflections up to a tenth of the NW length occurred. However, the measured strength did not correlate with this lateral deflection.

The bending experiments of sample A show a strength two times larger than the tensile experiments. Also, there is less scatter in the bending experiment data. This can be explained by the fact that only a small volume is highly stressed in the bent NW, so that it is less probable to encounter a structural defect that could initiate failure than in a tensile stressed NW, where the whole NW is highly stressed. 

The average Young's modulus of 97 GPa is 30\% lower than that of bulk (144 GPa along [0001] \cite{wern2004}). The most relevant systematic error we can think of is the measurement of the NW diameter. Because of the finite diameter of the SEM electron beam and the bright edges of the SEM images, the diameters might be measured too large. This could result in an underestimation of the Young's modulus, especially since the diameter enters squared in the calculation of $E$. The cantilevers spring constant was calculated by a finite element model including the cantilever holder and the tip, based on the dimensions measured in the SEM. We estimate the error on the spring constant to be lower than 15\%. The systematical error on strain can be neglected at it is a ration of two lenghts measured at the same magnification, only distortion of the SEM image could have an effect.

Physical reasons for a lower modulus could be vacancies present in the NWs. The effect of temperature on the Young's modulus can be neglected. Using the 1-dimensional heat equation and the heat conductivity of bulk ZnO, it can be shown that the temperature is not increased by more than 1 \textdegree C, assuming that all the energy of the electron beam is absorbed at the top of the NW and that no heat is extracted by the AFM tip, but the substrate is a perfect heat sink.

\section{Conclusions}

In conclusion, the mechanical strength of ZnO NWs was measured with a nanomanipulator inside a SEM. From bending experiments the fracture strain was 7.7\% and the tensile strength was 3.7 - 5.5 GPa for different samples, no plastic deformation was observed. The high strength indicate that no major structural defects are present in the NWs. From the tensile experiment the Young's modulus could be extracted to be about 100 GPa, that is 30\% lower than the bulk value and a factor of 2-3 higher than measured in bending experiments \cite{huang2006,bai2003,ni2006,song2005}. Possible reasons are: (1) differences in the measurement techniques. Our tensile approach is a direct one for measuring Young's modulus; (2) different dimensions and/or shape of the NWs which leads to different surface-to-volume ratios; (3) different level of vacancies within the NWs resulting from variance in their synthesis conditions.

\begin{acknowledgments}
The authors thank NanoWorld AG Neuch\^{a}tel for providing the AFM tips. S.H. thanks the Swiss National Science Foundation for financial support.
\end{acknowledgments}

\end{document}